\documentclass[twocolumn,showpacs,preprintnumbers,amsmath,amssymb]{revtex4-1}

\usepackage{graphicx}
\usepackage{dcolumn}
\usepackage{bm}

\begin{document}

\title{On the anomalous dynamics of capillary rise in porous media}

\author{Yulii D. Shikhmurzaev}
\email{Y.D.Shikhmurzaev@bham.ac.uk} \affiliation{School of
Mathematics, University of Birmingham, Birmingham,  B15 2TT, UK.}

\author{James E. Sprittles}
\email{sprittles@maths.ox.ac.uk} \affiliation{Mathematical
Institute, University of Oxford, Oxford OX1 3LB, UK.}

\date{\today}

\begin{abstract}
The anomalous dynamics of capillary rise in a porous medium
discovered experimentally more than a decade ago (Delker et al.,
Phys.\ Rev.\ Lett.\ 76 (1996) 2902) is described. The developed
theory is based on considering the principal modes of motion of
the menisci that collectively form the wetting front on the Darcy
scale. These modes, which include (i) dynamic wetting mode, (ii)
threshold mode and (iii) interface de-pinning process, are
incorporated into the  boundary conditions for the bulk equations
formulated in the regular framework of continuum mechanics of
porous media, thus allowing one to consider a general case of
three-dimensional flows. The developed theory makes it possible to
describe all regimes observed in the experiment, with the time
spanning more than four orders of magnitude, and highlights the
dominant physical mechanisms at different stages of the process.
\end{abstract}


\maketitle

\section{Introduction}

The propagation of the liquid-fluid interfaces through porous media
is central to a wide range of natural phenomena and industrial
applications, with the latter including enhanced oil recovery,
hydrogeology, fuel cells, carbon dioxide sequestration to mention
but a few.  This topic remains the subject of intensive research,
both experimental and theoretical, comprehensively reviewed in a
number of articles over the past forty years
\cite{Wooding-review:1976,Payatakes-review:1982,Brenner-review88,Olbricht-review-1996,Alava-review04}.
The main aspects of research have been the rate of propagation of
the wetting front
\cite{Labajos-Broncano-1999,Schoelkopf:2000,Gombia:2008,Reyssat:2009},
the wetting front's roughening \cite{Horvath:95,Soriano05} and
stability \cite{Tullis:2007}, as well as the related problems of the
formation and dynamics of the pockets of the displaced phase
(bubbles, ganglia) left behind the front
\cite{Payatakes-review:1982,Bernadiner-1998,Suekane:2010}. The first
of these aspects is of particular importance as it ultimately
determines the main macroscopic characteristics of the process in
many applications.

\begin{figure}
\centerline{\includegraphics[scale=0.5]{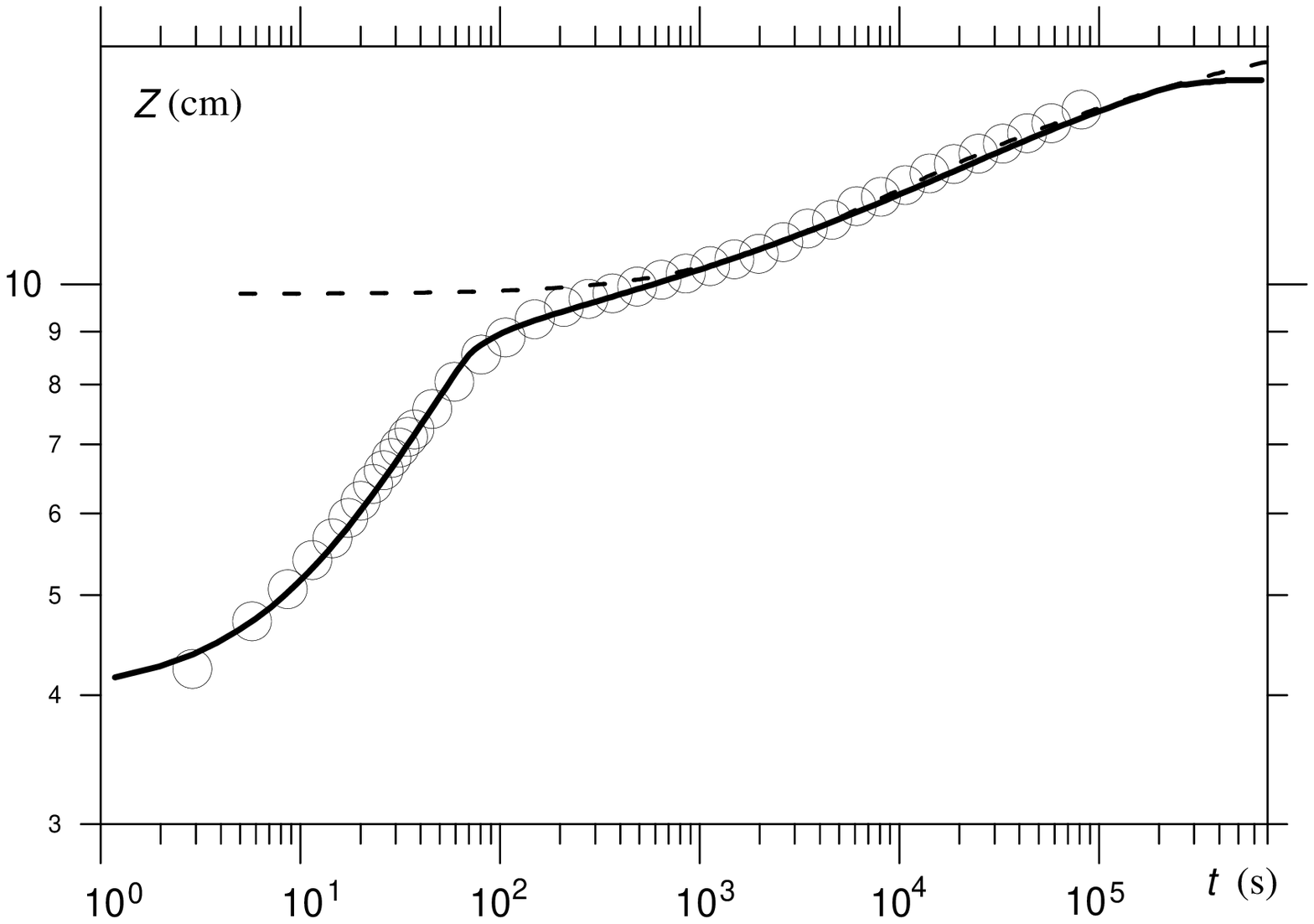}}
 \caption{Time-dependence of the imbibition height in 1D capillary rise.
  Circles: experimental data from \cite{Delker-etal-1996}
  for beads $253\ \mu$m in diameter;
  dashed line: fit considered in \cite{Delker-etal-1996};
  solid line: present theory.
  }
\label{fig:Delker96}
\end{figure}

As has been discovered experimentally more than a decade ago by
Delker and his co-workers \cite{Delker-etal-1996}, besides the
common situation where a wetting front propagates through a porous
medium broadly in accordance with Washburn's model
\cite{Washburn-1921}, which balances the driving force due to the
(presumed constant) capillary pressure of a meniscus and viscous
resistance as in the Poiseuille-type flow, for some media, such as
porous matrices made of packed spherical beads, the initial
Washburn-type imbibition is followed by a completely different and
in many ways `anomalous' regime. A representative set of data taken
from \cite{Delker-etal-1996} is shown in Fig.~\ref{fig:Delker96}.
Similar results have been reported later by Lago and Araujo
\cite{Lago-Araujo-2001}. The essence of the discovered effect is
that, if the height $h$ of the capillary rise (measured from some
initial level to remove from consideration the entrance effects) is
plotted against time $t$ on the $\log$-$\log$ scale
(Fig.~\ref{fig:Delker96}), one can immediately see two distinct
regions. Roughly speaking, for about two minutes the liquid climbs
2/3 of its eventual (maximum) height $h_{max}$ in the Washburn-like
regime after which it takes many hours for it to advances across the
remaining 1/3 of $h_{max}$, with the wetting front moving in
small-amplitude jumps on the pore scale \cite{Lago-Araujo-2001}. The
$\log$-$\log$ plot of this second regime shows a clear
concave-convex sequence which indicates that the dynamics there is
more complex than what one would expect from some power-law fit and
the accompanying arguments. Another intriguing feature of the
phenomenon is that $h_{max}$ is determined by the balance of
capillarity and gravity, i.e.\ the factors that, together with
viscous resistance, determine the dynamics of the Washburn regime,
although what looked like the Washburn regime has been abandoned
after a couple of minutes from the onset of the capillary rise and
for hours the process is distinctly non-Washburnian.

The experimental data has been discussed qualitatively in terms of
interface pinning and random capillary forces
\cite{Delker-etal-1996,Lago-Araujo-2001}, but on the quantitative
level the only outcome is that a simple equation
\begin{equation}
 \label{Delker-fit}
 dh/dt=v_0(F/F_T-1)^\beta
\end{equation}
expressing the rate of the capillary rise as a function of a
driving force $F$ and a threshold value $F_T$ leads to an
``anomalously large'' exponent $\beta$ \cite{Delker-etal-1996}, so
that $h$ deviates from the experimental data for small times and
unphysically diverges as time goes to infinity. The dashed line in
Fig.~\ref{fig:Delker96} corresponds to
\begin{equation}
\label{actual-fit}
h=H_c - (H_c-h_1)[1+A(t-t_1)]^{1/(1-\beta)}
\end{equation}
that has been deduced from (\ref{Delker-fit}) and used in
\cite{Delker-etal-1996}; the values of the constants $H_c$, $h_1$,
$A$, $t_1$ and $\beta$ are given in \cite{Delker-etal-1996}.
Although the fitting curve (\ref{actual-fit}) is able to describe
only a finite time span, the general ideas of the interface
pinning and random forces that might de-pin the interface and
allow it to move further seem fruitful, and a question that arises
naturally is how to embed them into the regular framework of
continuum mechanics of porous media, as opposed to just using {\it
ad hoc\/} semi-empirical equations for the wetting front evolution
in a one-dimensional flow.  Below, we address this question on the
basis of an earlier developed approach to the modelling of the
wetting front dynamics in porous media based on considering
different modes of motion that menisci go through on the pore
scale and the corresponding technique of conjugate problems
\cite{SS-11a}. Then, we will discuss how the experimental
phenomenon in question is seen through other modelling approaches.

\section{Macroscopic (Darcy-scale) description}

In the continuum framework, the wetting front $\partial\Omega_1$
is a moving boundary which, together with other boundaries
$\partial\Omega_2$, confines a domain $\Omega$ where the Darcy
equation,
\begin{equation}
\label{Darcy-1}
 \mathbf{u}=-(\kappa/\mu)\nabla (p+\rho gz),
 \qquad(\mathbf{r}\in\Omega),
\end{equation}
and the continuity equation, $\nabla\cdot\mathbf{u}=0$, for the
average velocity $\mathbf{u}$ and pressure $p$ operate.  Then, the
wetting front evolution is part of the solution of a properly
formulated problem for these bulk equations. The Darcy equation
(\ref{Darcy-1}) is written in the form already accounting for
gravity with $\rho$ being the density of the liquid, $g$ the
gravitational acceleration, $z$ the coordinate directed against
gravity, $\kappa$ the permeability of the porous matrix and $\mu$
the liquid's viscosity; the coordinates are represented in terms
of the position vector $\mathbf{r}$; hereafter the pressure is
measured with respect to the (presumed constant) pressure in the
displaced gas.

An appropriate starting point for the modelling is the recently
developed approach \cite{SS-11a} that gives boundary conditions
for Laplace's equation for $p$,
\begin{equation}
\label{Laplace-p}
 \nabla^2p=0,\qquad(\mathbf{r}\in\Omega),
\end{equation}
which follows from the Darcy and continuity equations above and
can be used to replace the latter. The key idea of this approach
is to consider the {\it modes of motion\/} which the menisci that
collectively form the wetting front undergo as the wetting front
propagates. The simplest model formulated in the framework of this
approach accounts for the two main modes: (i) the wetting mode,
where, on the pore scale, the contact line moves forward,
essentially in the Washburn regime but accounting for a {\it
dynamic}, i.e.\ velocity-dependent, contact angle $\theta_d$, and
(ii) the threshold mode, where the contact line gets pinned and
the meniscus bends as the pressure on it increases until the
contact angle reaches some threshold value $\theta_*$ when the
meniscus can go back into the wetting mode with the contact line
moving again. This increase of pressure on the meniscus as the
contact line gets pinned is similar to what one would have on a
piston sucking a liquid into a pipe if the motion of the piston is
blocked. For the porous medium, the maximum possible pressure on
the meniscus in the threshold mode, $\bar{p}|_{\partial\Omega_1}$,
is the solution of a {\it conjugate\/} problem \cite{SS-11a}
\begin{equation}
\label{conjugate}
 \nabla^2\bar{p}=0,
 \quad
 (\mathbf{r}\in\Omega);
 \qquad
 \mathbf{n}\cdot\nabla(\bar{p}+\rho gz)|_{\partial\Omega_1}=0,
\end{equation}
with the boundary condition for $\bar{p}$ on $\partial\Omega_2$
being the same as for $p$; $\mathbf{n}$ is the outward normal to
$\partial\Omega_1$. Thus, the idea of the interface pinning is
already in the model, used in formulating the boundary conditions
on the wetting front that we will recapitulate below, and in this
work we consider how the model can be generalized to incorporate
the idea of random forces that could lead to de-pinning of the
interface and describe the observed features of the phenomenon
mentioned earlier.

On the moving wetting front the kinematic and dynamic boundary
conditions for (\ref{Laplace-p}) have the form
\begin{equation}
\label{kinematics-general}
 \frac{\partial f}{\partial t} + \mathbf{u}\cdot\nabla f=0,
\end{equation}
\begin{equation}
 \label{p=}
 p=A_1 p_1+A_2 p_2,
 \quad (\mathbf{r}\in\partial\Omega_1),
\end{equation}
where for a one-dimensional capillary rise $f(\mathbf{r},t)\equiv
z-h(t)$, $p_1$, $p_2$ are the averaged pressures and $A_1$, $A_2$
are the spatio-temporally averaged fractions of the unit area of
the free surface corresponding to the two modes of motion
($A_1+A_2=1$). For the wetting mode one has
\begin{equation}
\label{p_1=}
 p_1=-2\sigma\cos\theta_d/a,
\end{equation}
where $\sigma$ is the liquid-gas surface tension, $a$ is the
effective radius of the capillary and the dependence of the dynamic
contact angle $\theta_d$ on the meniscus speed $u_1$ is given by
\cite{SS-11a,TheBook}
\begin{equation}
\label{theta_d=}
 \frac{u_1}{U_{cl}}=
 \left(
 \frac{
 (1+(1-\rho^s_{1e})\cos\theta_s)(\cos\theta_s-\cos\theta_d)^2}
 {4(\cos\theta_s+B)(\cos\theta_d+B)}
 \right)^{1/2},
\end{equation}
where $B=(1-\rho^s_{1e})^{-1}(1+\rho^s_{1e}u_0(\theta_d))$,
$\theta_s$ is the static contact angle,
$$
 u_0(\theta_d)=\frac{\sin\theta_d-\theta_d\cos\theta_d}
 {\sin\theta_d\cos\theta_d-\theta_d},
 \qquad
 U_{cl}
 =\left(\frac{\gamma\rho^s_0(1+4\alpha\beta)}{\tau\beta}\right)^{1/2}
$$
is the characteristic speed associated with the parameters that the
`additional' physics of wetting brings in to resolve the well-known
`moving contact-line problem'
\citep{Dussan-review79,TheBook,dry_facts}, $\rho^s_0$,
$\rho^s_{1e}$, $\alpha$, $\beta$, $\gamma$, $\tau$ are material
constants characterizing the contacting media whose values can be
found elsewhere \citep{TheBook,Blake-me02}.

In the threshold mode, the contact line gets pinned and the
meniscus, experiencing an increase in pressure on it, bends so that
the contact angle varies from $\theta_d$, i.e.\ the value with which
the meniscus arrives at the threshold mode, to $\theta_*$, which is
the value at which the meniscus `breaks through' the threshold and
the process goes back to the wetting mode. From the dynamics of this
type of motion, under the assumption that the meniscus retains the
shape of a spherical cap with its radius varying as the meniscus
bends, one has that \cite{SS-11a}
\begin{equation}
 \label{u_2=}
 u_2=aJ(\theta_d)T^{-1},
\end{equation}
\begin{equation}
 \label{p_2=}
 p_2=P - \frac{2\sigma}{u_1T}
 \left(\frac{Pa}{2\sigma}+\cos\theta_d\right)
 J(\theta_d),
\end{equation}
where
\begin{equation}
 \label{big_P=}
 P = \bar{p}|_{\partial\Omega_1},
\end{equation}
$$
J(\theta_d)=
\left[\frac{1}{2}\tan\left(\frac{\theta}{2}-\frac{\pi}{4}\right)
+\frac{1}{6}\tan^3\left(\frac{\theta}{2}-\frac{\pi}{4}\right)
\right]_{\theta_d}^{\theta_*},
$$
$[f]_a^b\equiv f(b)-f(a)$, and the time $T$ that the meniscus
spends in the threshold mode is given by
\begin{equation}
 \label{old_T=}
 T=T_2\left(\theta_d,\frac{Pa}{2\sigma}\right)
 \equiv a u_1^{-1}\left(\frac{Pa}{2\sigma}
+\cos\theta_d\right)
I\left(\theta_d,\frac{Pa}{2\sigma}\right),
\end{equation}
$$
I\left(\theta_d,\frac{Pa}{2\sigma}\right)
=\int_{\theta_d}^{\theta_*} \frac{d\theta}
{(1+\sin\theta)^2(Pa/(2\sigma)+\cos\theta)}.
$$
For a one-dimensional capillary rise, as follows from
(\ref{conjugate}), the stagnation pressure
$\bar{p}|_{\partial\Omega_1}$ is given simply by
$\bar{p}|_{\partial\Omega_1}=p_0-\rho gh(t)$, where $p_0$ is the
prescribed pressure at $z=0$.

Finally, the velocity of the wetting front as a whole,
$u_n=\mathbf{n}\cdot\mathbf{u}|_{\partial\Omega_1}$ is given by an
expression
\begin{equation}
 \label{u_n=}
 u_n=A_1u_1+A_2u_2,
\end{equation}
which is similar to (\ref{p=}). For the menisci intermittently going
through the wetting and threshold modes as the wetting front
propagates, the coefficients $A_1$ and $A_2$ can be viewed as
reflecting the fraction of the time spent in each mode as the
meniscus travels over the length of averaging that introduces the
Darcy scale (or, equivalently, the fraction of the interfacial area
corresponding to each mode of motion over the time interval of
averaging that introduces the Darcy time scale), i.e.\ the
spatio-temporal averages we mentioned earlier. They are given by
\cite{SS-11a}
\begin{equation}
\label{alphas}
 A_1=\frac{s_1u_2}{s_2u_1+s_1u_2},
\quad A_2=\frac{s_2u_1}{s_2u_1+s_1u_2},
\end{equation}
where
\begin{equation}
\label{s=}
 s_1(\theta_d,\theta_*)=
 \left\{
 \begin{array}{ll}
 1, & \theta_d-\theta_*\ge0\\
 s_{10}, & \theta_d-\theta_*<0
 \end{array}
 \right.,
 \qquad
 s_2=1-s_1,
\end{equation}
and $s_{10}$ $(<1)$ is a characteristic of the porous matrix. Then,
as should be expected, the slowest (controlling) mode of motion
makes a greater contribution to the average pressure and velocity at
the wetting front, and if the velocity $u_i$ corresponding to the
$i$th mode reaches zero, one will have $A_i=1$ and the pressure at
the wetting front, that is now at rest, will become equal to $p_i$.
Thus, the wetting front will stop propagating in two cases: (a)
$u_1=0$ and hence $\theta_d=\theta_s$, which means that the meniscus
has reached its equilibrium state corresponding to the maximum
imbibition height $h_{max}$, and (b) $u_2=0$ so that the wetting
front still has a capacity to propagate further but the contact line
became pinned (threshold mode) and the pressure that mounts on the
meniscus in this case, even when it reaches its maximum possible
value $\bar{p}|_{\partial\Omega_1}$, is insufficient to push the
meniscus through. Mathematically, in the last case we have that
$\bar{p}|_{\partial\Omega_1}$, which goes down as $h$ increases,
becomes equal to $p_*=-2\sigma\cos\theta_*/a$ and hence is unable to
make the contact angle greater than $\theta_*$, which would allow
the meniscus to resume its motion in the wetting mode: in this case
$I(\theta_d,Pa/(2\sigma))$ and hence the time $T$ go to infinity.
The height corresponding to this last case depends only on
$\theta_*$, and the meniscus reaches it in a finite time
\cite{SS-11a}.

\section{Subcritical interface de-pinning}

It is convenient to introduce the `stagnation' contact angle
$\bar{\theta}$ corresponding to the stagnation pressure
$\bar{p}|_{\partial\Omega_1}$ by
$\bar{\theta}=\arccos(-\bar{p}|_{\partial\Omega_1}a/(2\sigma))$.
Then, if $\bar{\theta}>\theta_s$, the wetting front has the
potential to propagate but if at the same time
$\bar{\theta}\le\theta_*$ the stagnation pressure
$\bar{p}|_{\partial\Omega_1}$ is unable to push the meniscus through
the threshold mode. In a sense, $\theta_*-\bar{\theta}$ can be
viewed as a quantitative measure of the potential barrier that has
to be overcome to get the meniscus back into the wetting mode when
$\bar{p}|_{\partial\Omega_1}$ is `subcritical', i.e.\ less than
$p_*$, and hence unable to push the meniscus through the threshold.

Importantly, since, until the wetting front reaches its equilibrium
position at the maximum height, at every moment individual menisci
are not in the same mode of motion (and, for the threshold motion,
not even in the same stage of it), the Darcy-scale pressures we are
considering, including the stagnation pressure
$\bar{p}|_{\partial\Omega_1}$, represent average values, whereas on
the pore scale one also has pressure fluctuations due to mutual
influences of menisci. These fluctuations are unimportant when the
stagnation pressure $\bar{p}|_{\partial\Omega_1}$ is capable of
pushing the meniscus through the threshold mode. However, as
$\bar{p}|_{\partial\Omega_1}$ goes down to $p_*$, the time $T$
needed to overcome the threshold increases, so that, when it becomes
large enough, it is the fluctuations that increasingly become the
mechanism of de-pinning, and when $\bar{\theta}\le\theta_*$ it is
only the random fluctuations that can de-pin the menisci.

The simplest way of accounting for the subcritical de-pinning due to
random fluctuations as this mechanism takes over from the `regular'
de-pinning due to the stagnation pressure is to assume that, once
$T_2$ becomes greater than a certain value $T_+$, it is these random
factors that will de-pin the interface and determine the time it
stays in the threshold mode. For the `regular' stagnation pressure
$\bar{p}|_{\partial\Omega_1}$ one has that $T_2\to\infty$ as
$\bar{p}|_{\partial\Omega_1}\to p_*$ and
$\bar{p}|_{\partial\Omega_1}$ is no longer able to push the meniscus
through when $\bar{p}|_{\partial\Omega_1}<p_*$. The probability of
the random factors de-pinning the interface should be expected to go
down as $p_*-\bar{p}|_{\partial\Omega_1}$ (or equivalently
$\theta_*-\bar{\theta}$) increases. The simplest way of generalizing
the model to incorporate the above scenario mathematically is to
replace (\ref{old_T=}) and (\ref{big_P=}) respectively with
\begin{equation}
 \label{new_T=}
 T=\left\{
 \begin{array}{ll}
 \displaystyle
 T_2\left(\theta_d,\frac{a\bar{p}|_{\partial\Omega_1}}{2\sigma}\right),
 & \hbox{if }
   T_2(\bar{p}|_{\partial\Omega_1})\le T_+\\
 T_++k(\theta_+-\bar{\theta})^2, & \hbox{if }
   T_2(\bar{p}|_{\partial\Omega_1})> T_+
 \end{array}
 \right.,
\end{equation}
\begin{equation}
 \label{new_big_P=}
 P=\left\{
 \begin{array}{ll}
 \bar{p}|_{\partial\Omega_1}, & \hbox{if }
   T_2(\bar{p}|_{\partial\Omega_1})\le T_+\\
 p_+, & \hbox{if }
   T_2(\bar{p}|_{\partial\Omega_1})> T_+
 \end{array}
 \right.,
\end{equation}
where $p_+$ is determined by $T_2(p_+)=T_+$, and
$\theta_+=\arccos(-p_+a/(2\sigma))$. Since $T_2$ rises steeply
only when $\bar{\theta}$ is close to $\theta_*$, in practice one
has that $\theta_+\approx\theta_*$ and $p_+\approx p_*$.

Now, we have a closed model, which, unlike {\it ad hoc\/} formulae
for one-dimensional propagation of the wetting front, is applicable
for a general case of three-dimensional flows.  In order to describe
a particular flow involving a moving wetting front, one has to solve
Laplace's equation (\ref{Laplace-p}) for $p$ in $\Omega$ whose
boundary $\partial\Omega_1$ evolves according to
(\ref{kinematics-general}), where $\mathbf{u}$ is given by
(\ref{Darcy-1}), subject to the dynamic condition (\ref{p=}), where
equations (\ref{p_1=})--(\ref{p_2=}),
(\ref{u_n=})--(\ref{new_big_P=}) close the formulation and
$\bar{p}|_{\partial\Omega_1}$ is the solution of the conjugate
problem (\ref{conjugate}); the boundary conditions on
$\partial\Omega_2$ for both $p$ and $\bar{p}$ are the same and,
together with the initial shape of $\Omega$, they specify a
particular problem.

\begin{figure}
\centerline{\includegraphics[scale=0.5]{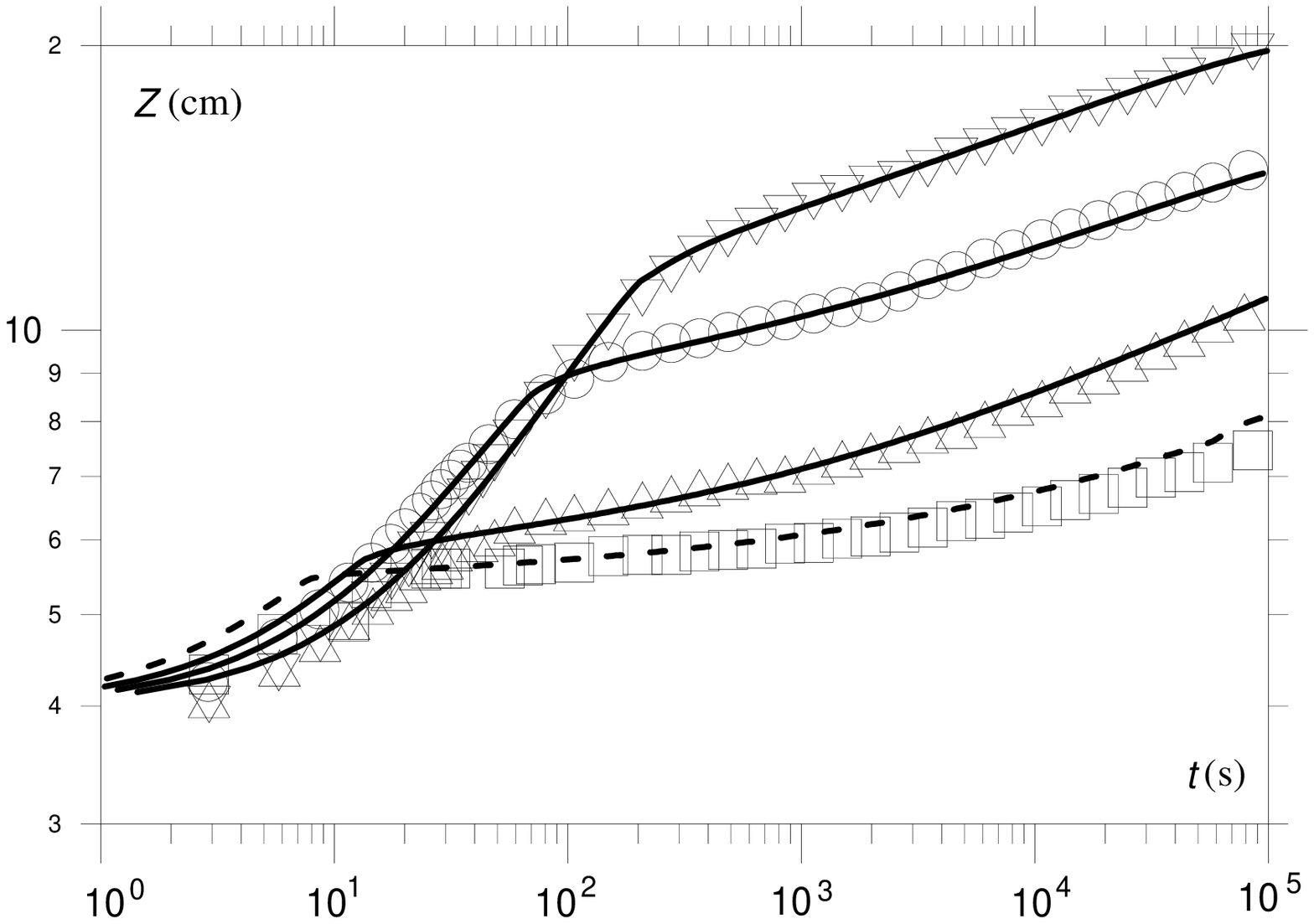}}
 \caption{All four experimental sets from \cite{Delker-etal-1996}
  for beads' diameters
 $180$~$\mu$m ($\triangledown$),
  $253$~$\mu$m ($\circ$),
  $359$~$\mu$m ($\vartriangle$), and
  $510$~$\mu$m ($\square$).
  The corresponding theoretical curves are given as solid lines for
  the first 3 sets of data and as a dashed line for the last.
  }
\label{fig:all_four}
\end{figure}

In the case of a one-dimensional capillary rise Laplace's
equations for $p$ and $\bar{p}$ give that these are linear
functions of $z$, and equations (\ref{Darcy-1}),
(\ref{Laplace-p}), (\ref{kinematics-general}) yield
$$
\frac{dh}{dt}=\frac{\kappa}{\mu}\left(
 \frac{p_0-p(h,t)}{h} - \rho g\right),
$$
which together with the algebraic equations (\ref{p=}), where
$p|_{\partial\Omega_1}\equiv p(h,t)$, (\ref{p_1=})--(\ref{p_2=}),
(\ref{u_n=}), where $u_n=dh/dt$,
(\ref{alphas})--(\ref{new_big_P=}), with
$\bar{p}|_{\partial\Omega_1}=p_0-\rho gh$ as the solution of the
conjugate problem, form a closed system for $h$, $p(h,t)$, $p_1$,
$\theta_d$, $u_1$, $u_2$, $p_2$, $A_1$, $A_2$, $s_1$ and $s_2$.
The results of comparing the numerical solution corresponding to
the experimental flow conditions of \cite{Delker-etal-1996} with
the data are shown in Fig.~\ref{fig:Delker96}. As one can see, the
solid curve representing the computed solution describes the data
very well over the whole time period spanning more than four
orders of magnitude. The theoretical curve also levels off as
$t\to\infty$, indicating that the capillary rise does eventually
come to a halt. Comparison of the theory with all four sets of
experimental data from \cite{Delker-etal-1996} is shown in
Fig.~\ref{fig:all_four}. (The dashed line in this figure is used
to indicate that for the beads' diameter of 510~$\mu$m, strictly
speaking, the theory is used outside its limits of applicability
as the whole advancement of the wetting front is less than 40
beads' diameters, so that it is difficult to talk about the
separation of scales required for the continuum mechanics approach
to work. Indeed, for this approach to be applicable, there should
exist an intermediate scale much larger than the pore size and at
the same time much smaller than the macroscopic length scale on
which the flow is described. In the case of 510~$\mu$m beads,
`much large' and `much smaller' would mean 6 times larger or
smaller, which is clearly not sufficient to ensure acceptable
accuracy.)

It is noteworthy that, although the initial regime where the curve
in Fig.~\ref{fig:Delker96} rises steeply looks Washburn-like, it
actually involves both the wetting and the threshold modes of
motion. As in \cite{SS-11a}, the presence of the
threshold-overcoming motion becomes pronounced only when $h$
climbs close to $h_*=(p_0-p_*)/(\rho g)$, i.e.\ when $\bar{p}$
becomes close to $p_*$ or, in other words, $\bar{\theta}$ close to
$\theta_*$. For the results presented in Fig.~\ref{fig:Delker96}
$\theta_*=67^\circ$, $\theta_s=0^\circ$, $s_{10}=0.7$, $\mu
U_{cl}/(\kappa\rho g)=10^2$,  $\rho^s_{1e}=0.6$, $T_+/T_0=3$,
$k/T_0=4\times10^3$, where $T_0=2\sigma\mu/((\rho g)^2a\kappa)$,
and it is $\theta_s$, $\theta_*$, $T_+/T_0$ and $k/T_0$ that are
most important. Since  in the experiment, as described in
\cite{Delker-etal-1996}, the bottom of the test section was
located approximately 4~cm above the bottom of the porous sample,
we have to start the calculations from the bottom of the sample
with $p_0=\rho g Z_{ini}$, $Z_{ini}=4$~cm and, to compare theory
with the data, measure the time from the moment the wetting front
reaches $Z_{ini}$. To describe the data in
Fig.~\ref{fig:all_four}, only a variation of $\theta_*$ and
$K=k/T_0$ is required: $\theta_{*180}=64^\circ$,
$\theta_{*253}=67^\circ$, $\theta_{*359}=78^\circ$;
$K_{359}=3K_{253}=12K_{180}$., $K_{180}=10^3$.

It is worth mentioning that, although in the model we represent
capillary effects in the pores using the capillary pressure and
viscous resistance corresponding to an `effective' circular
cross-section, whereas in the experiment with the porous matrix
made of spherical beads neither the `chambers' nor the `throats'
of the porous medium had circular cross-sections, no adjustment of
the results was required: we used the radius of the bead as $a$ in
the model with no subsequent calibration of the time and length
scales. This shows that a `representative' way of modelling the
porous medium, as opposed to much more difficult way of
incorporating into a model the exact porous structure determined
via elaborate and expensive experiments, allows one to incorporate
all the main features of the process on the pore scale, including
the actual physics of dynamic wetting, and obtain good results for
the flow on the Darcy scale.

\section{Discussion}

It is instructive to look at the obtained theoretical result and the
experiment it addresses from the viewpoint of the different
modelling approaches used to describe two-phase flows in porous
media.  These approaches broadly fall into two general classes,
`representative' and `simulative'. They are not antagonistic as, in
theory, if the same pore-scale physics and the same characteristics
of the porous medium are accounted for in the models formulated in
the framework of each of these approaches, then the results produced
by these models are expected to converge and describe the same
macroscopic behaviour of the wetting front.

The present model has been developed in the framework of the
`representative' approach, where the equations and boundary
conditions are formulated on the Darcy scale, i.e.\ the scale
implying that the continuum limit has already been taken, and the
properties of the porous matrix are `represented' in terms of the
permeability coefficient (or tensor, if the porous medium is
anisotropic), effective size of the pores (or the corresponding
distribution), effective threshold angles, etc. Importantly, since
the pore-scale wetting process is modelled realistically, with the
velocity (as well as material) dependence of the dynamic contact
angle, the model captures naturally that the wetting front has the
capacity to move forward when the contact angle is greater than the
static value $\theta_s$, i.e.\ when the interface has not reached
its maximum height determined by the balance of capillary and
gravity forces. In other words, as the meniscus gets de-pinned, the
fact that the contact angle differs from its equilibrium value
$\theta_s$ and that the dynamic contact angle is velocity-dependent
act as a mechanism that moves the interface until the dynamic
contact angle goes down to $\theta_s$ and the interface reaches its
maximum height.

The subcritical de-pinning mechanism that comes into action when
$\bar{p}|_{\partial\Omega_1}\le p_*$ is formulated implicitly, in
terms of the `potential barrier' $\theta_+-\bar{\theta}$ and the
`waiting time' $T_++k(\theta_+-\bar{\theta})^2$ required for the
random fluctuations to overcome it. Both of these characteristics
are Darcy-scale parameters. A natural way to develop the model
further would be to remove the direct link between the `potential
barrier' and the `waiting time' and instead explicitly introduce the
field of pressure fluctuations depending on the flow rate and
properties of the porous matrix. Then, this pressure fluctuation
field becomes an addition to $\bar{p}|_{\partial\Omega_1}$ as a
breakthrough factor. For this explicitly introduced mechanism, the
results obtained in the present work would serve as a guideline,
indicating one of the outcomes that this mechanism should produce.

The implicit mechanism of the subcritical interface de-pinning on
the Darcy scale that we have introduced can be viewed as a
macroscopic manifestation of the dynamics of avalanches
\cite{Dougherty-Carle:1998}. Avalanches qualitatively stem from the
same physics as the one considered here and, in a sense, can be
regarded as a medium-scale phenomenon, i.e.\ between the pore scale
and the Darcy scale. The idea of linking the Darcy-scale subcritical
interface de-pinning  and the dynamics of avalanches agrees with the
fact that the avalanche-type events have been observed
experimentally by Lago and Araujo \cite{Lago-Araujo-2001} in the
anomalous regime of the wetting front propagation. The avalanche
behaviour of the wetting fronts is currently being investigated in
terms of scaling laws \citep{Rost_etal:2007}, and the present
theory, with its explicit accounting for different pressures
corresponding to different modes of motion and their spatio-temporal
weights, offers a macroscopic framework for the mathematical
description of these medium-scale events.  Then, the subcritical
interface de-pinning in the anomalous regime of the wetting front
propagation could be viewed as the macroscopic outcome of a
succession of avalanches with decreasing probabilities.

The `simulative' approach to the modelling of two-phase flows in
porous media is based on replacing the actual porous matrix with a
regular network of `chambers' and capillary `throats' connecting
them
\cite{Fatt:1956-164,Lenormand88,Aker00,Joekar-Niasar-2010,Markicevic:2011}.
In order to partially compensate the anisotropy inherent in this
approach where, as is the case in most works, the chambers are
placed at the nodes of a regular lattice, the sizes of both
chambers and throats are made random following certain prescribed
distributions. The macroscopic (Darcy-scale) characteristics of
the process are obtained as a result of the appropriate averaging.
The simulative approach has the appeal of what looks like a
numerical experiment offering a transparent link between the
pore-scale and the Darcy-scale dynamics, but, unlike the case of
molecular-dynamics simulations with regard to macroscopic fluid
mechanics, this appeal is moderated by a number of factors,
notably the fact that the actual dynamics on the pore scale is not
computed. Instead, it is essentially represented in terms of a
Washburn-type dynamics, thus by-passing the `moving contact-line
problem' \cite{Dussan-review79} and the associated physics of
dynamic wetting \cite{dry_facts,Blake-review:2006}. It is also
important to note that the rigid geometric structure of the
network simulating the actual porous matrix imposes unavoidable
limitations on the macroscopic transport properties of the porous
medium that the network is purports to simulate. By contrast, the
representative approach can introduce any tensorial and
topological characteristics of the porous medium that the
corresponding model requires. Having in mind the above
shortcomings of the representative approach, it is interesting to
look at what it can produce with regard to the anomalous regime of
the wetting front propagation.

Bijeljic {\it et al.} \cite{Markicevic:2011} performed the network
modelling of the capillary rise and compared their results to the
truncated set of data taken from Lago and Araujo
\cite{Lago-Araujo-2001}, with the original experiment by Delker et
al.\ \cite{Delker-etal-1996} mentioned but not used for comparison
with the simulations. The simulations agree well with the
experimental data for the times $t\le4\times10^4$~s after which the
simulated height levels off.  However, when one takes the full set
of experimental data reported by Lago and Araujo, i.e. for the times
up to $t=2\times10^5$~s, one can immediately see that the wetting
front continues to climb. It is also worth noting that the simulated
height-vs-time curve in the $\log$-$\log$ coordinates is convex,
with the slope monotonically decreasing, whereas, as one can see in
Fig.~\ref{fig:Delker96}, the experimental data show a distinct
concave-convex sequence, i.e.\ after an initial decrease as the
anomalous regime is entered the slope picks up again until, finally,
the data start to level off, asymptotically approaching the maximum
height.  The same trend was observed earlier by Diggins {\it et al.}
\cite{Diggins-Ralston-1990} whose data are given in Fig.~12 of Lago
and Araujo \cite{Lago-Araujo-2001}.  This figure in
\cite{Lago-Araujo-2001} also clearly shows that, although the
time-dependence of the height of the capillary rise following from
the Washburn-type interplay of capillarity, viscosity and gravity
can describe the `regular' regime {\it and\/} be fitted to the
initial stage of the `anomalous' regime, it is nowhere near a
satisfactory description of the latter once the full set of data is
considered, as the height-vs-time curve in the logarithmic
coordinates picks up again and climbs much higher than the above
fitting predicts.

From the viewpoint of the theory developed in the present work, the
main deficiency of the currently implemented network models is that
they essentially deal only with one --- wetting --- mode of motion
of the menisci. Then, setting aside the minor (in comparison with
the effects considered here) variations introduced by randomizing
the size distributions, these models broadly reproduce the
Washburn-type dynamics of the wetting front. As a result, once
gravity starts to balance capillarity as the driving force, the
wetting front slows down and comes to a halt. Essentially, the
fitting of the simulated curve to the experimental data for the very
beginning of the anomalous regime is produced by adjusting the
maximum height of the capillary rise, whereas, as we pointed out in
the introduction, the intriguing feature of the anomalous regime is
precisely the fact that it lies in between the `normal' Washburnian
regime of the imbibition and the maximum height that is also
Washburnian.

The capillary network approach can be modified in a relatively
straightforward way to account for the threshold mode of motion. The
main element in this modification should be equipping the `chambers'
with threshold characteristics, such as the pressure required to
overcome the threshold, which, besides material properties, can
depend on the number of menisci reaching the same chamber. The
implementation of the subcritical de-pinning is more challenging as
this would require accounting for fluctuations of pressure
experienced by the liquid, i.e.\ replacing the Washburn-type models
of the flow in the `throats' by an essentially unsteady motion that
takes into account the unsteady processes in the neighboring
chambers and throats. An intermediate check for such a model could
be its ability to produce avalanches as the medium-scale phenomena
that on the Darcy scale result in the anomalous regime of
imbibition.

\section{Conclusion}

The developed theory shows that the new approach to the modelling
of the propagation of wetting fronts in porous media based on
considering specific modes of motion that the menisci of the pore
scale undergo as the front propagates allows one to incorporate
critical phenomena and adequately describe experimental data for
the anomalous regime of imbibition. Accounting for the random
pore-scale forces macroscopically, in terms of the `potential
barriers' and the corresponding times required for the random
forces to overcome these barriers, allowed the simplest model
formulated in the framework of the new approach to describe the
whole experimental curve, from the Washburn regime to the (also
Washburn) maximum imbibition height with the anomalous regime in
between. The proposed theory could be used as a guide for the
porous network modelling and the study of the anomalous imbibition
regime as the manifestation of the dynamics of avalanches.

This publication was based on work supported in part by Award No.\
KUK-C1-013-04 , made by King Abdullah University of Science and
Technology (KAUST).

\bibliographystyle{abbrvnat}
\bibliography{references} 

\end{document}